\documentstyle[12pt,epsf]{article}
\begin{document}

\title{\large \bf
To the problem of electron-phonon interaction and "d-wave pairing"
in high-$T_c$ oxides.}
\author{\normalsize L.S. Mazov \\
\normalsize \it Institute for Physics of Microstructures,
\normalsize \it Russian Academy of Science\\
\normalsize \it Nizhny Novgorod, GSP-105, 603600, Russia }
\date{}

\maketitle
{\scriptsize It is supported
a recent proposal by Maksimov \cite{Max} that
electron-phonon interaction (EPI) (see, e.g. Gruneisen-Bloch formulae)
determines the linearity of temperature dependence of resistivity for both
HTSC-cuprates and most of metals at $T \ge \Theta_{D}/5$ (normal state
for HTSC). However, it is here emphasized that resistivity is not
proportional to $T$ in this temperature range but is only linear in $T$.
This fact indicates to temperature-independent contribution to
normal-state resistivity in HTSC-cuprates (magnetic contribution in our
treatment) which behavior, in its turn, indicates to possible magnetic
(SDW-like) phase transition before SC transition in HTSC-system.
SDW-gap (measured as pseudogap with d-wave symmetry, in our
treatment) is formed at symmetrical parts of the Fermi surface above
$T_c$ and persists in SC state. The magnitude of SDW-gap is essentially
larger than that for SC-gap. So, SDW formed can mimic d-wave symmetry of
the order parameter measured below $T_c$ (which is now attributed to
d-wave pairing). On the other hand, since Gruneisen-Bloch curve appears
to be the geometric locus for the onset of SC transition at different
magnetic fields (as in LTSC), this fact may be considered as evidence for
EPI-nature (BSC) of SC transition (s-wave) in high-$T_c$-oxides.
Also, there is presented a proposal to model stripe structure (SDW/CDW)
of $CuO_2$-plane in HTSC-cuprates using both static (subnanotechnology)
and dynamic (solid-state plasma physics) methodics to increase $T_c$.}
\\ [1cm]
{\sf Submitted to Usp.Fiz.Nauk as a Comment to the
paper by E.Maksimov \cite{Max} (February, 2001)} \\ [1cm]

In a recently published paper by Maksimov \cite{Max} it was noted that
linearity of temperature dependence of resistivity $\rho(T)$, observed in
a normal state of high-$T_c$ oxides, is also characteristic for most
metals ({\it magnetic} ones are included) at $T \ge \Theta_{D}/5$ (fig.2
in \cite{Max}). In this temperature region resistivity of the metal is
due to phonon scattering of charge carriers ($\rho(T) = \rho_{ph}(T)$)
and is well described by Gruneisen-Bloch formulae (here $\Theta_D$ is a
Debye temperature of corresponding metal). In this connection, it may be
noted the following. The problem of relation of this linearity
to copper oxides begin discuss, in fact, just after
discovery of high-$T_c$ superconductivity (see, e.g. well-known review by
Gor'kov and Kopnin \cite{GK}). However, already first calculations by
Allen, Pickett and Krakauer \cite{APK} have demonstrated that
phonon scattering (calculated for parameters of high-$T_c$ oxides) though
provides linearity of $\rho_{ph}(T)$-dependence in the normal state of
HTSC but corresponding dependences lay essentially lower than
experimental ones. Moreover, same conclusion follows
from Gruneisen-Bloch formulae too: even from fig.2 in \cite{Max}(see,
also the book by Kittel "Introduction to Solid-State Physics" (John Wiley
and Sons, Inc.) or the monograph by Ziman "Electrons and Phonons"
(Oxford at the Clarendon Press, 1960 ))
it's seen that in this temperature range $\rho_{ph}(T)$ is not
proportional to $T$  $(\rho_{ph}\sim T)$ as it is usually considered but
is only linear in $T$, cutting (at extrapolation) negative part at the
ordinate axis which value is comparable with that of $\rho_{ph}(T)$ at $T
\ge \Theta_D/5$. At the same time, in HTSC cuprates linear dependence
corresponding to $\rho(T)$ in the normal state comes through the
coordinate origin ("in best single-crystalline samples") or usually
cuts positive part at the $\rho$-axis. This fact indicates to
presence in HTSC system of additional (to phonon) contribution $\rho_m(T)$
caused by scattering mechanism which scattering rate only slightly
depends on temperature ($\rho_m(T) \approx const$) in this temperature
range. It seems to be natural to attract here the magnetic
scattering mechanism: parent compounds $La_2CuO_4$ and $YBa_2Cu_3O_6$
are antiferro(AF)-magnetic insulators and, at their doping, in HTSC
system there persist AF spin fluctuations which are an
effective origin of scattering for charge carriers in the normal
state. The same conclusion (as one of the possibilities ) was in fact
made by authors of \cite{APK}, too. Analysis of the HTSC behavior
performed by us in the beginning of 90's \cite{M1}, in frames of
the model \cite{Vons} ($\rho_{tot} = \rho_{ph} + \rho_m$, where
$\rho_m$ is a magnetic contribution in resistivity of HTSC) used
at analysis of resistivity $\rho(T)$ in magnetic metals,
permits one to describe the "anomalous" behavior of HTSC cuprates
at ($H,T$)-plane, in details. As it's seen from fig.1 (see \cite{M2})
Gruneisen-Bloch curve (dashed curve in the fig.1) separates from
experimental $\rho(T)$-curves (1-4) a contribution $\rho_m(T)$ (shaded
area in fig.1a). The behavior of so-extracted magnetic contribution
$\rho_m(T)$ (fig.1b) evidents (cf. with \cite{Vons}) about phase
transition in magnetic subsystem of HTSC cuprates from spin-disordered
($\rho_m \approx const$ in the normal state) to magnetically-ordered
state ($\rho_m = 0$), modulated magnetic structure (like to spin
density wave (AF SDW)) coexisting with superconductivity at $T \le
T_k(H)$. As it has been followed from analysis of available in that time
literature data (see, \cite{M1}) SDW in the system is realized in the
form of quasistatic stripe structure in $CuO_2$-plane (see, insert in
fig.1b) when magnetic stripes (triangles) alternate with charge
stripes (circles) (charge density wave (CDW)) (see, also the review by
Bianconi \cite{Bia}). It should be underlined that parts of
resistive curves which lay above points of intersection with
Gruneisen-Bloch curve ($T \ge T_k(H)$) relate to magnetic (SDW) phase
transition, i.e. correspond to normal rather than resistive state as
it is usually considered (in other words, above $T = T_k(H)$ in HTSC
sample there are no SC vortices, at all).

Furthermore, in the intersection points of (experimental) resistive
curves with Gruneisen-Bloch one it appears the shoulder which, in fact,
indicate to the end of magnetic (SDW) phase transition $T_k(H) =
T_{SDW}^{order}(H)$ (see above) and the onset ($T_k(H) = T_c^{onset}(H)$)
of superconducting transition itself. It should be underlined that even
at $H = 26 T$ the onset point of SC transition also lays at
Gruneisen-Bloch curve (though
this curve already deviates from linearity, dashed
curve in fig.1c) which fact evidents about applicability of
Gruneisen-Bloch formulae to analysis of HTSC-cuprates system at $T <
\Theta_D/5$, too.  Moreover, the fact that Gruneisen-Bloch curve (see
fig.1c) is, in fact, a geometric locus for the onset of SC transition
at different magnetic fields $T_c^{onset}(H)$ (as in LTSC) may be
considered as evidence for EPI-nature (BCS) of SC transition (s-wave)
in high-$T_c$ oxides. As additional evidence for such conclusion may be
also considered the linearity of temperature dependence of upper critical
magnetic field $H_{c2}(T)$ (with $H_{c2}(0) \approx 45 T$ and $\xi_{ab}
\approx 25 \AA$) characteristic for GL and BCS theory in the nearity of
$T_c(0)$ (see, insert in fig.1c). The dependence of $H_{c2}(T)$ is formed
from points $T_c^{onset}(H)$ at the ($H,T$)-plane \cite{M1}. (It
should be noted that in literature there is significant spread in
values of $H_{c2}(0)$ up to now, and moreover, correct resistive methodics
to determine $T_c(H)$ in HTSC cuprates is absent, at all).

The picture obtained was not unexpected: the problem of coexistence of
superconductivity and magnetism was firstly considered yet
in 1956, by Ginzburg
\cite{Ginz1}, and then was developed by many researchers (see, e.g.
review by Machida \cite{Mach}). From the theory by Machida for the system
of itinerant electrons with interplay between superconductivity and
magnetism it follows that with decreasing temperature it is firstly formed
SDW-gap at symmetrical parts of the Fermi surface (normal state), and
only then, SC-gap. Such anisotropy of SDW-gap is in agreement with
$d_{x^2 -y^2}$-wave symmetry of pseudogap which appearance (together with
stripe structure) in the normal state of HTSC-cuprates
(at $T = T^* = T_{SDW}^{onset}$, in
given model) is now observed in many experiments.

Then, same conclusion about SDW/CDW-origin for the pseudogap was made
by Klemm \cite{Kl} from Argonne National Lab., USA, on the basis
of detailed analysis of direct measurements of energy gap characteristics
in HTSC-cuprates.
Since, pseudogap also persists in SC state, then its d-wave symmetry
mimics the symmetry of the order parameter measured in SC state (e.g.,
with using of ARPES-methodics or tunnel experiments).

So, evidence obtained by us that SC transition with phonon mechanism in
HTSC cuprates occurs only after formation of additional (SDW) order
in the normal state, is in agreement with
conclusions of \cite{Max,Ginz2} that "in cuprates ... there are
essential an electron-phonon interaction (EPI) as well as ... spin
interaction" \cite{Ginz2} which, in fact, is namely "something
determining, together with EPI, SC mechanism in HTSC systems" \cite{Max}.
Analysis performed in frames of above approach \cite{M1,M2} permits one
significantly clear the picture of "anomalous" behavior of HTSC-cuprates
and answer two main questions arising when describing the phenomena in
frames of pure EPI: "observed anisotropic d-pairing" (or more correctly,
strong anisotropy of energy gap) and " large value of
$2\Delta(0)/k_BT_c"$ \cite{Max,Ginz2}. These features now attributed
to $d_{x^2 -y^2}$-wave superconductivity, appear to be, in fact,
direct consequence of coexistence of
SDW and SC order parameters below $T =
T_k(H)$ (see, fig.1a) when SDW-gap with its $d_{x^2 -y^2}$-wave symmetry
and relatively large magnitude ($\Delta_{SC} < \Delta_{SDW}$ \cite{Mach})
prevails in the electron energy spectra \cite{M1,Kl,M2} measured in
experiments. As for experiment, then investigations performed in last
decade with both conventional and new (fast and local (ARPES, EXAFS,
STM)) methodics led to quantitative agreement with detailed picture
obtained in \cite{M1} (for more details, see \cite{M2}). So, in recent
experiments on elastic neutron scattering in La-based cuprates it was
observed that superconductivity and static magnetic order appear at the
same temperature (see, above: $T_{SDW}^{order}(0) = T_c^{onset}(0)$,
fig.1a). In addition, the model (SDW/CDW) here considered permits one to
propose \cite{M2} both static (with layer-by-layer sputtering (cf.
\cite{Bia,Ginz2})) and dynamic (the creation of standing wave)
methods to increase $T_c$ (for detailed discussion of the problem, see
monograph \cite{HTS}) in such systems. As first example of using (in
fact) this mechanism (SDW/CDW/SC) may be considered recent communication
in Nature ($\bf 394$, 453, 1998) about twofold increase of $T_c$
in nanostructure with
using of thin film of $La$-based cuprate on the substrate with slightly
different lattice periods (for more details, see \cite{Bia,M2}).


\newpage

\begin{figure}
\epsfysize=7in
\epsfbox{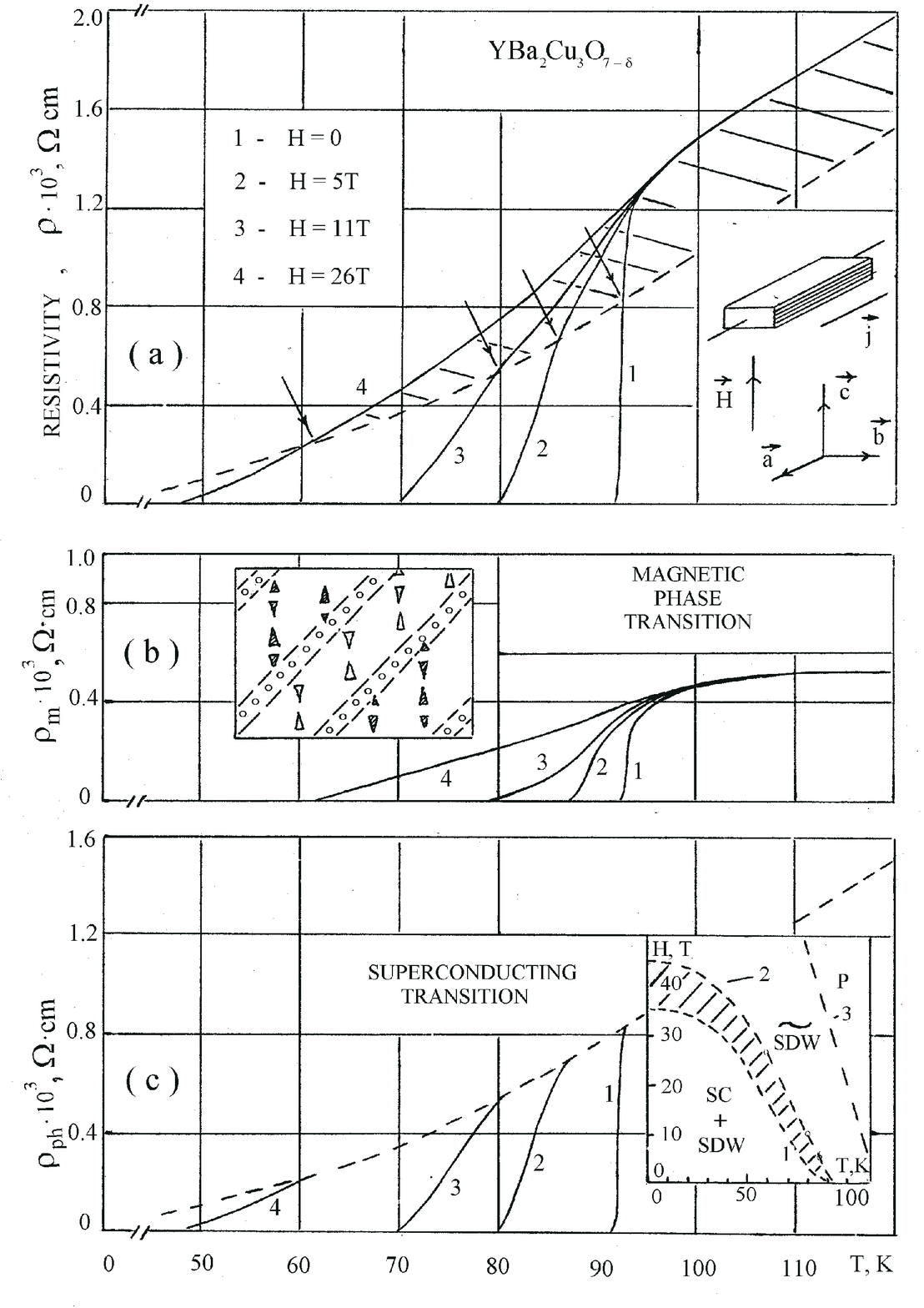}
\caption{Temperature dependence of in-plane resistivity in YBCO 
single crystal (see, \cite {M2}).}
\end{figure}

\end{document}